\documentclass[12pt,journal,final,onecolumn]{IEEEtran}

\usepackage[dvips]{graphicx}
\usepackage{epsfig}
\usepackage[cmex10]{amsmath}
\usepackage{amssymb}
\usepackage{amsthm}
\usepackage{amsfonts}
\usepackage{bm}
\usepackage{cite}
\usepackage[tight,footnotesize]{subfigure}
\usepackage{algorithm}
\usepackage{algpseudocode}

\graphicspath{{figs/}}

\input{niesen.def}

\begin{document}

\title{Decentralized Coded Caching Attains\\ Order-Optimal Memory-Rate Tradeoff}

\author{Mohammad Ali Maddah-Ali and  Urs Niesen%
\thanks{The authors are with Bell Labs, Alcatel-Lucent. Emails:
\{mohammadali.maddah-ali, urs.niesen\}@alcatel-lucent.com}
\thanks{The material in this paper has been presented in part at the 51st Annual Allerton Conference on Communication, Control, and Computing, Oct. 2013.}%
}

\maketitle

\begin{abstract} 
    Replicating or caching popular content in memories distributed
    across the network is a technique to reduce peak network loads.
    Conventionally, the main performance gain of this caching was
    thought to result from making part of the requested data available
    closer to end users. Instead, we recently showed that a much more
    significant gain can be achieved by using caches to create
    coded-multicasting opportunities, even for users with different
    demands, through coding across data streams. These
    coded-multicasting opportunities are enabled by careful content
    overlap at the various caches in the network, created by a central
    coordinating server. 
    
    In many scenarios, such a central coordinating server may not be
    available, raising the question if this multicasting gain can still
    be achieved in a more decentralized setting. In this paper, we
    propose an efficient caching scheme, in which the content placement
    is performed in a decentralized manner. In other words, no
    coordination is required for the content placement. Despite this
    lack of coordination, the proposed scheme is nevertheless able to
    create coded-multicasting opportunities and achieves a rate close to
    the optimal centralized scheme.
\end{abstract}

\section{Introduction}
\label{sec:intro}

Traffic in content delivery networks exhibits strong temporal
variability, resulting in congestion during peak hours and resource
underutilization during off-peak hours. It is therefore desirable to try
to ``shift'' some of the traffic from peak to off-peak hours.  One
approach to achieve this is to exploit idle network resources to
duplicate some of the content in memories distributed across the
network. This duplication of content is called content placement or
caching. The duplicated content can then be used during peak hours to
reduce network congestion.

From the above description, it is apparent that the network operates in
two different phases: a content placement phase and a content delivery
phase. In the placement phase, the network is not congested, and the
system is constrained mainly by the size of the cache memories. In
the delivery phase, the network is congested, and the system is
constrained mainly by the rate required to serve the content requested
by the users. The goal is thus to design the placement phase such that
the rate in the delivery phase is minimized.

There are two fundamentally different approaches, based on two distinct
understandings of the role of caching, for how the placement and the
delivery phases are performed.
 
\begin{itemize}
    \item \emph{Providing Content Locally:} In the first, conventional,
        caching approach, replication is used to make part of the
        requested content available close to the end users. If a user
        finds part of a requested file in a close-by cache memory, that
        part can be served locally. The central content server only
        sends the remaining file parts using simple orthogonal unicast
        transmissions. If more than one user requests the same file,
        then the server has the option to multicast a single stream to
        those users. 
        
        Extensive research has been done on this conventional caching
        approach, mainly on how to exploit differing file popularities
        to maximize the caching
        gain~\cite{dowdy82,almeroth96,dan96,korupolu99,meyerson01,baev08,borst10}.
        The gain of this approach is proportional to the fraction of the
        popular content that can be stored locally. As a result, this
        conventional caching approach is effective whenever the local
        cache memory is large enough to store a significant fraction of
        the total popular content.
    \item \emph{Creating Simultaneous Coded-Multicasting Opportunities:}
        In this approach, which we recently proposed
        in~\cite{maddah-ali12a}, content is placed in order to allow the
        central server to satisfy the requests of several users with
        \emph{different} demands with a \emph{single} multicast stream.
        The multicast streams are generated by coding across the
        different files requested by the users. Each user exploits the
        content stored in the local cache memory to enable decoding of
        its requested file from these data streams. Since the content
        placement is performed before the actual user demands are known,
        it has to be designed carefully such that these
        coded-multicasting opportunities are available simultaneously
        for all possible requests. 

        In~\cite{maddah-ali12a}, we show that this simultaneous
        coded-multicasting gain can significantly reduce network
        congestion. Moreover, for many situations, this approach results
        in a much larger caching gain than the one obtained from the
        conventional caching approach discussed above. Unlike the
        conventional approach, the simultaneous coded-multicast approach
        is effective whenever the aggregate \emph{global} cache size
        (i.e., the cumulative cache available at all users) is large
        enough compared to the total amount of popular content, even
        though there is no cooperation among the caches.
\end{itemize}

As mentioned above, the scheme proposed in~\cite{maddah-ali12a}, relies
on a carefully designed placement phase in order to create
coded-multicasting opportunities among users with different demands. A
central server arranges the caches such that every subset of the cache
memories shares a specific part of the content. It is this carefully
arranged overlap among the cache memories that guarantees the
coded-multicasting opportunities simultaneously for all possible user
demands. 

While the assumption of a centrally coordinated placement phase was
helpful to establish the new caching approach in~\cite{maddah-ali12a},
it limits its applicability. For example, the identity or even just the
number of active users in the delivery phase may not be known several
hours in advance during the placement phase. As another example, in some
cases the placement phase could be performed in one network, say a WiFi
network, to reduce congestion in the delivery phase in another network,
say a cellular network. In either case, coordination in the placement
phase may not be possible. 

This raises the important question whether lack of coordination in the
placement phase eliminates the significant rate reduction promised by the
simultaneous coded-multicast approach proposed in~\cite{maddah-ali12a}.
Put differently, the question is if simultaneous coded-multicasting
opportunities can still be created without a centrally coordinated
placement phase.

In this paper, we answer this question in the positive by developing a
caching algorithm that creates simultaneous coded-multicasting
opportunities without coordination in the placement phase. More
precisely, the proposed algorithm is able to operate in the placement
phase with an unknown number of users situated in isolated networks and
acting independently from each other. Thus, the placement phase of the
proposed algorithm is \emph{decentralized}. In the delivery phase, some
of these users are connected to a server through a shared bottleneck
link. In this phase, the server is first informed about the set of
active users, their cache contents, and their requests. The proposed
algorithm efficiently exploits the multicasting opportunities created
during the placement phase in order to minimize the rate over the shared
bottleneck link.  We show that our proposed decentralized algorithm can
significantly improve upon the conventional uncoded scheme.  Moreover,
we show that the performance of the proposed decentralized coded caching
scheme is close to the performance of the centralized coded scheme
of~\cite{maddah-ali12a}. 

\begin{figure}[htbp]
    \centering
    \hspace{-1.5cm}\includegraphics{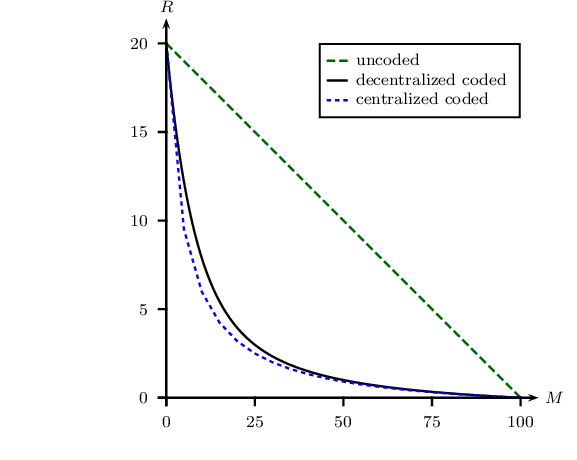}
    \caption{Performance of different caching schemes for a system with
        $20$ users connected to a server storing $100$ files through a
        shared bottleneck link.  The horizontal axis is the size of the
        cache memory (normalized by the file size) at each user; the
        vertical axis shows the peak rate (again normalized by the file
        size) over the shared link in the delivery phase. The dashed
        green curve depicts the rate achieved by the conventional
        uncoded caching scheme advocated in the prior literature. The
        solid black curve depicts the rate achieved by the decentralized
        coded caching scheme proposed in this paper. The dashed blue
        curve depicts the rate achieved by the centralized coded caching
        algorithm from the recent paper~\cite{maddah-ali12a}.}
    \label{fig:intro}
\end{figure}

These two claims are illustrated in Fig.~\ref{fig:intro} for a system
with $20$ users and $100$ pieces of content. For example, when each user
is able to cache $25$ of the files, the peak rate of the conventional
uncoded scheme is equivalent to transmitting $15$ files. However, in the
proposed decentralized coded scheme, the peak rate is equivalent to
transmitting only about $3$ files. By comparing this to the performance
of the centralized coded scheme, we can see that the rate penalty for
decentralization of the placement phase of the caching system is modest.

The remainder of this paper is organized as follows.
Section~\ref{sec:problem} formally introduces the problem setting.
Section~\ref{sec:main_proposed} presents the proposed algorithm.  In
Section~\ref{sec:performance}, the performance of the proposed algorithm
is evaluated and compared with the uncoded and the centralized coded
caching schemes. In Section~\ref{sec:extensions}, the results are
extended to other topologies of practical interest. Section
\ref{sec:discussions} discusses various implications of the results.

\section{Problem Setting}
\label{sec:problem}

To gain insight into how to optimally operate content-distribution
systems, we introduce here a basic model for such systems capturing the
fundamental challenges, tensions, and tradeoffs in the caching
problem. For the sake of clarity, we initially study the
problem under some simplifying assumptions, which will be relaxed later,
as is discussed in detail in Sections~\ref{sec:extensions}
and~\ref{sec:discussions}.

\begin{figure}[htbp]
    \centering
    \includegraphics{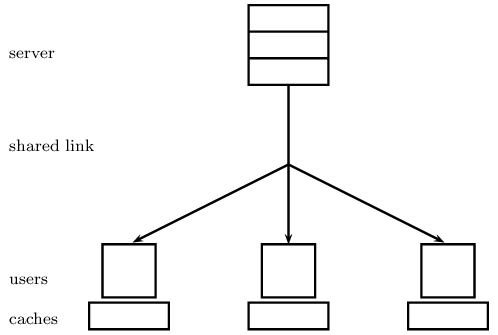}
    \caption{Caching system considered in this paper. A
    server containing $N$ files of size $F$ bits each is connected
    through a shared link to $K$ users each with a cache of size $MF$
    bits. In the figure, $N=K=3$ and $M=1$. } 
    \label{fig:setting}
\end{figure}

We consider a content-distribution system consisting of a server
connected through an error-free\footnote{Any errors in this link have
presumably been already taken care of using error correction coding.}
shared (bottleneck) link to $K$ users. The server stores $N$ files each
of size $F$ bits. The users each have access to a cache able to store
$MF$ bits for $M\in[0,N]$. This scenario is illustrated in
Fig.~\ref{fig:setting}.

The system operates in two phases: a \emph{placement} phase and a
\emph{delivery} phase. The placement phase occurs when the network load
is low. During this time, the shared link can be utilized to fill the
caches of the users. The main constraint in this phase is the size of
the cache memory at each user. The delivery phase occurs after the
placement phase when the network load is high.  At this time, each user
requests one file from the server, which proceeds to transmit its
response over the shared link. Given the output of the shared link
(observed by all users) and its cache content, each user should be able
to recover its requested file. The main constraint in this phase is the
load of the shared link. The objective is to minimize the worst-case
(over all possible requests) load of the shared link in the delivery
phase.

We now formalize this problem description. In the placement phase,  each
user is able to fill its cache as an arbitrary function (linear,
nonlinear, \ldots) of the $N$ files subject only to its memory
constraint of $MF$ bits with $M\in[0,N]$. We emphasize that the requests
of the users are not known during the placement phase, and hence the
caching function is not allowed to depend on them.

In the delivery phase, each of the $K$ users requests one of the $N$
files and communicates this request to the server. Let
$d_k\in\{1,\dots,N\}$ be the request of user $k\in\{1,\dots,K\}$. The
server replies to these requests by sending a message over the shared
link, which is observed by all the $K$ users.  Let
$R^{(d_1,\dots,d_K)}F$ be the number of bits in the message sent by the
server. We impose that each user is able to recover its requested file
from the content of its cache and the message received over the shared
link with probability arbitrary close to one for large enough file size
$F$. Denote by 
\begin{equation*}
    R \defeq \max_{d_1,\dots,d_K}R^{(d_1,\dots,d_K)}
\end{equation*}
the worst-case normalized rate for a caching scheme.

Our objective is to minimize the rate $R$ in order to minimize the
worst-case network load $RF$ during the delivery phase. Clearly, $R$ is
a function of the cache size $MF$. In order to emphasize this
dependence, we will usually write the rate as $R(M)$.  The function
$R(M)$ expresses the \emph{memory-rate tradeoff} of the
content-distribution system. 

The following example illustrates the definitions and notations and
introduces the uncoded caching approach advocated in most of the prior
literature. This uncoded caching scheme will be used as a benchmark
throughout the paper. 

\begin{example}[\emph{Uncoded Caching}]
    \label{eg:conventional}    
    Consider the caching problem with $N=2$ files and $K=2$ users each
    with a cache of size $M=1$.  In the uncoded caching scheme, each of
    the two files $A$ and $B$ are split  into two parts of equal size,
    namely $A=(A_1,A_2)$ and $B=(B_1,B_2)$.  In the placement phase,
    both users cache $(A_1,B_1)$, i.e., the first part of each file.
    Since each of these parts has size $F/2$, this satisfies the memory
    constraint of $MF=F$ bits. 

    Consider now the delivery phase of the system. Assume that each user
    requests the same file $A$, i.e., $d_1=d_2=1$. The server responds
    by sending the file part $A_2$ of size $F/2$ bits. Clearly, from
    their cache content and the message received over the shared link,
    each user can recover the requested file $A=(A_1,A_2)$. The
    (normalized) rate in the delivery phase is $R^{(1,1)} = 1/2$.

    Assume instead that user one requests file $A$ and user two requests
    file $B$, i.e., $d_1=1$ and $d_2=2$. The server needs to transmit
    $(A_2,B_2)$ of size $F$ bits to satisfy these requests, resulting in
    a rate in the delivery phase of $R^{(1,2)} = 1$.  It is easy to see
    that this is the worst-case request, and hence $R = 1$ for this
    scheme. 

    For general $N$, $K$, and $M$, the uncoded scheme caches
    the first $M/N$ fraction of each of the $N$ files. Therefore, in the
    delivery phase, the server has to send the remaining  $1-M/N$
    fraction of the requested files.  The resulting rate in the delivery
    phase, denoted by $R_U(M)$ for future reference, is 
    \begin{equation*}
        R_U(M) \defeq K\cdot(1-M/N)\cdot\min\bigl\{1, N/K \bigr\}.
    \end{equation*}
    For $N=K=2$ and $M=1$, this yields $R_U(1) = 1$, as before.

    As we will see, this conventional caching scheme can be
    significantly improved upon. In particular, see
    Example~\ref{eg:illustration} in Section~\ref{sec:main_proposed}.
\end{example}

One important feature of the uncoded scheme introduced in
Example~\ref{eg:conventional} is that it has a \emph{decentralized}
placement phase. By that we mean that the cache of each user is filled
independently of other users. In particular, the placement operation of
a given user neither depends on the identity nor the number of other
users in the system. As a result, the users could, in fact, contact
different servers at different times for the placement phase. Having a
decentralized placement phase is thus an important robustness property
for a caching system. This is discussed further in
Sections~\ref{sec:main_proposed} and \ref{sec:extensions}.

As was mentioned earlier, the system description introduced in this
section makes certain simplifying assumptions. In particular, we assume
a system having a single shared broadcast link, with a cache at each
user, and we focus on worst-case demands, with synchronized user
requests in the delivery phase.  All these assumptions can be relaxed,
as is discussed in Sections~\ref{sec:extensions} and
\ref{sec:discussions}.

\section{A Decentralized Coded Caching Algorithm}
\label{sec:main_proposed}

We now present a new algorithm (referred to as decentralized coded
caching in the following) for the caching problem.  In the statement of
the algorithm, we use the notation $V_{k,\mc{S}}$ to denote the bits of
the file $d_k$ requested by user $k$ cached exclusively at users in
$\mc{S}$. In other words, a bit of file $d_k$ is in $V_{k,\mc{S}}$ if it
is present in the cache of every user in $\mc{S}$ and if it is absent
from the cache of every user outside $\mc{S}$. We also use the notation
$[K] \defeq \{1,2,\dots,K\}$ and $[N] \defeq \{1,2,\dots,N\}$. 

The proposed algorithm consists of a placement procedure and two
delivery procedures. In the placement phase, we always use the same
placement procedure. In the delivery phase, the server chooses the
delivery procedure minimizing the resulting rate over the shared link.

\begin{algorithm}[h!]
    \caption{Decentralized Coded Caching}
    \label{alg:1}
    \begin{algorithmic}[1]
        \Procedure{Placement}{}
        \For{$k\in[K], n\in[N]$}
        \State user $k$ independently caches a subset of $\tfrac{MF}{N}$
        bits of file $n$, chosen uniformly at random \label{alg:1_cache}
        \EndFor
        \EndProcedure
        \Statex
        \Procedure{Delivery}{$d_1,\dots,d_K$}
        \For{$s=K, K-1, \ldots, 1$} \label{alg:1_sloop}
        \For{$\mc{S}\subset[K]: \card{\mc{S}}=s$} \label{alg:1_Sloop}
        \State server sends \(\oplus_{k\in\mc{S}} V_{k,\mc{S}\setminus\{k\}}\) \label{alg:1_send}
        \EndFor 
        \EndFor 
        \EndProcedure
        \Statex
        \Procedure{Delivery'}{$d_1,\dots,d_K$}
        \For{$n\in[N]$}
        \State server sends enough random linear combinations of bits in file
        $n$ for all users requesting it to decode \label{alg:1_send2}
        \EndFor
        \EndProcedure
    \end{algorithmic}
\end{algorithm}

\begin{remark}
    \label{rem:alg}
    The $\oplus$ operation in Line~\ref{alg:1_send} of
    Algorithm~\ref{alg:1} represents the bit-wise XOR operation.  All
    elements \(V_{k,\mc{S}\setminus\{k\}}\) are assumed to be zero
    padded to the length of the longest element.
\end{remark}

We illustrate the Algorithm~\ref{alg:1} with a small example.
\begin{example}[\emph{Decentralized Coded Caching}]
    \label{eg:illustration}
    Consider the caching problem with $N=2$ files $A$ and $B$, and $K=2$
    users each with a cache of size $M\in[0,2]$. In the placement phase
    of Algorithm~\ref{alg:1}, each user caches a subset of $MF/2$ bits
    of each file independently at random, satisfying the memory
    constraint. As a result, each bit of a file is cached by a specific
    user with probability $M/2$.  
    
    Let us focus on file $A$. The actions of the placement procedure
    effectively partition file $A$ into $4$ subfiles, 
    \begin{equation*}
        A=(A_\emptyset, A_{1},  A_{2}, A_{1,2}),
    \end{equation*}
    where, for $\mc{S}\subset \{1,2\}$, $A_{\mc{S}}$ denotes the bits of
    file $A$ that are stored in the cache memories of users in
    $\mc{S}$. For example, $A_{1,2}$ are the bits of $A$
    available in the cache memories of users one and two, whereas $A_1$
    are the bits of $A$ available exclusively in the cache memory of
    user one.\footnote{To avoid heavy notation, we write $A_{1,2}$ as
    shorthand for \(A_{\{1,2\}}\). Similarly, we write $V_{1,2}$ for
    \(V_{1,\{2\}}\).}
    
    By the law of large numbers,
    \begin{equation*}
        \card{A_{\mc{S}}} 
        \approx (M/2)^{\card{\mc{S}}}(1-M/2)^{2-\card{\mc{S}}}F
    \end{equation*}
    with probability approaching one for large enough file size $F$.
    Therefore, we have with high probability:
    \begin{itemize}
        \item $\card{A_\emptyset}/F$ is approximately $(1-M/2)^2$.
        \item $\card{A_{1}}/F$
            and $\card{A_{2}}/F$ are approximately $(M/2)(1-M/2)$.
        \item $\card{A_{1,2}}/F$ is approximately $(M/2)^2$.
    \end{itemize}
    The same analysis holds for file $B$.  
    
    We now consider the delivery phase in Algorithm~\ref{alg:1}.  As we
    will see later (see Remark~\ref{rem:thm} below), for the scenario at
    hand the first delivery procedure will be used. Assume that user one
    requests file $A$ and user two requests file $B$. 
    
    The iteration in Line~\ref{alg:1_sloop} of Algorithm~\ref{alg:1}
    starts with $s=2$. By Line~\ref{alg:1_Sloop}, this implies that we
    consider the set $\mc{S} = \{1,2\}$. Observe that:
    \begin{itemize}
        \item The cache of user two contains $A_{2}$, which
            is needed by user one. Hence, \(V_{1,2}=A_{2}\).
        \item The cache of users one contains $B_{1}$, which
            is needed by user two. Hence, \(V_{2,1}=B_{1}\).
    \end{itemize}
    As a result, in Line~\ref{alg:1_send} of Algorithm~\ref{alg:1}, the
    server transmits $A_{2} \oplus B_{1}$ over the shared link. User one
    can solve for $A_{2}$ from the received message $A_{2} \oplus B_{1}$
    and the cached subfile $B_{1}$. User two can solve for $B_{1}$ from
    the message $A_{2} \oplus B_{1}$ and the cached subfile $A_{2}$.
    Therefore, $A_{2} \oplus B_{1}$ is simultaneously useful for $s=2$
    users.  Thus, even though the two users request different files, the
    server can successfully multicast useful information to both of
    them. We note that the normalized (by $F$) size of $A_{2} \oplus
    B_{1}$ is $(M/2)(1-M/2)$. 

    The second iteration in Line~\ref{alg:1_sloop} is for $s=1$. In this
    iteration, the server simply sends $V_{1,\emptyset}=A_\emptyset$ and
    $V_{2,\emptyset}=B_\emptyset$ in Line~\ref{alg:1_Sloop}. Each of
    these transmissions is useful for $s=1$ user and has normalized size
    $(1-M/2)^2$. 

    From $A_{2}$ computed in iteration one, $A_\emptyset$ received in
    iteration two, and its cache content $(A_1, A_{1,2})$, user one can
    recover the requested file $A=(A_\emptyset, A_{1},  A_{2},
    A_{1,2})$. Similarly, user two can recover the requested file $B$. 

    Summing up the contributions for $s=2$ and $s=1$, the aggregate size
    (normalized by $F$) of the messages sent by the server is
    \begin{equation*}
        (M/2)(1-M/2)+2(1-M/2)^2.
    \end{equation*}
    This can be rewritten as 
    \begin{equation*}
        2\cdot (1-M/2)\cdot \frac{1}{M}\bigl( 1-(1-M/2)^2 \bigr).
    \end{equation*}
    In particular, for $M=1$, the rate of Algorithm~\ref{alg:1} is
    $3/4$.
    
    This compares to a rate of $R_U(1) = 1$ achieved by the uncoded
    caching scheme described in Example~\ref{eg:conventional} in
    Section~\ref{sec:problem}. While the improvement in this scenario
    is relatively small, as we will see shortly, for larger values of
    $N$ and $K$, this improvement over the uncoded scheme can be
    large.
\end{example}

\begin{remark}[\emph{Unknown Number of Users during Placement Phase}]
    \label{rem:knownK}
    The placement procedure of Algorithm~\ref{alg:1} is
    \emph{decentralized}, in the sense that the user's caches are filled
    independently of each other. This implies that neither the identity
    nor even the number of users sharing the same bottleneck link during
    the delivery phase need to be known during the earlier placement
    phase. 

    This decentralization of the placement phase enables the
    content-distribution system to be much more flexible than a
    centralized placement phase. This flexibility is essential.  For
    example, in wireline networks, some of the users may not request any
    file in the delivery phase. In wireless networks, users may move
    from one network or cell to another, and hence might not even be
    connected to the same server in the two phases. In either case, the
    result is that the precise number and identity of users in the
    delivery phase is unknown in the placement phase.  One of the
    salient features of the decentralized algorithm proposed in this
    paper is that it can easily deal with these situations. 
    
    This flexibility is also crucial to deal with asynchronous user
    requests, as is explained in detail in
    Section~\ref{sec:extensions_synchronization}. It is also a key
    ingredient to extending the coded-caching approach to scenarios with
    nonuniform demands or with online cache updates, as is discussed
    further in Section~\ref{sec:discussions} and in the follow-up works
    \cite{niesen13} and \cite{pedarsani13}.
\end{remark}

\begin{remark}[\emph{Greedy Coding Strategy}]
    The first delivery procedure in Algorithm~\ref{alg:1} follows a
    \emph{greedy} strategy.  It first identifies and forms coded
    messages that are useful for all $s=K$ users. In the next iteration,
    it forms coded messages that are useful for subsets of $s=K-1$
    users.  The iteration continues until it identifies messages that
    are useful for only $s=1$ user. 
\end{remark}

\begin{remark}[\emph{Simplified Decision Rule}]
    \label{Remark:M}
    Algorithm~\ref{alg:1} provides two delivery procedures. The general
    rule is to choose the procedure which minimizes the resulting rate
    over the shared link.  A simple alternative rule to decide between
    these two procedures is as follows: if $M >1$, employ the first
    procedure; otherwise, employ the second procedure.  The performance
    loss due to this simpler rule can be shown to be small.\footnote{In
    fact, the achievable rate with this simpler rule is still within
    a constant factor of the optimal centralized memory-rate
    tradeoff. This follows from the proof of
    Theorem~\ref{thm:comparison_converse} with some minor
    modifications.}
\end{remark}

\begin{remark}[\emph{Knowledge of Cache Contents}]
    The delivery procedure in Algorithm~\ref{alg:1} assumes that the server
    knows which bits are cached at each user. In practice, each user
    will choose which bits to cache using a random number generator. By
    communicating only the seed value of this random number generator
    from the user back to the server, the server can reconstruct the
    cache contents of the user.
\end{remark}

\section{Performance Analysis}
\label{sec:performance}

We now analyze the performance of the proposed decentralized coded
caching scheme given by Algorithm~\ref{alg:1}.
Section~\ref{sec:performance_rate} provides an analytic expression for
the rate of Algorithm~\ref{alg:1}.
Section~\ref{sec:performance_decentralized} compares the proposed
decentralized coded caching scheme with the decentralized uncoded
caching scheme from Example~\ref{eg:conventional} (the best previously
known decentralized caching scheme).
Section~\ref{sec:performance_centralized} compares the proposed
decentralized coded caching scheme with the optimal centralized caching
scheme and the caching scheme from~\cite{maddah-ali12a} (the best known
centralized caching scheme).

\subsection{Rate of Decentralized Coded Caching Scheme}
\label{sec:performance_rate}

The performance of decentralized coded caching is analyzed in the next
theorem, whose proof can be found in
Appendix~\ref{sec:proofs_achievable1}. 

\begin{theorem}
    \label{thm:achievable1}
    Consider the caching problem with $N$ files each of size $F$ bits
    and with $K$ users each having access to a cache of size $MF$ bits
    with $M\in(0,N]$. Algorithm~\ref{alg:1} is correct and, for $F$
    large enough, achieves rate arbitrarily close to
    \begin{align*} 
        R_D(M) 
        & \defeq K\cdot(1-M/N)\cdot\min\biggl\{ 
            \frac{N}{KM}\bigl(1-(1-M/N)^K\bigr), \, \frac{N}{K} 
        \biggr\}.
    \end{align*}
\end{theorem}

\begin{remark}
    \label{rem:thm}
    We note that if $N\geq K$ or $M \geq 1$, then the minimum in
    $R_D(M)$ is achieved by the first term so that
    \begin{equation*} 
        R_D(M) = K\cdot(1-M/N)\cdot \frac{N}{KM}\bigl(1-(1-M/N)^K\bigr).
    \end{equation*}
    This is the rate of the first delivery procedure in
    Algorithm~\ref{alg:1}. Since $N\geq K$ or $M\geq 1$ is the regime of
    most interest, the majority of the discussion in the following
    focuses on this case.
\end{remark}

\begin{remark}
    \label{rem:zero}
    Theorem~\ref{thm:achievable1} is only stated for $M > 0$. For $M=0$
    Algorithm~\ref{alg:1} is easily seen to achieve a rate of 
    \begin{equation*}
        R_D(0) \defeq\min\{N,K\}.
    \end{equation*}
    We see that $R_D(0)$ is the continuous extension of $R_D(M)$ for $M
    >0$. To simplify the exposition, we will not treat the case $M=0$
    separately in the following.
\end{remark}

The rate $R_D(M)$ of Algorithm~\ref{alg:1} consists of three distinct
factors. The first factor is $K$; this is the rate without caching. The
second factor is $(1-M/N)$; this is a \emph{local} caching gain that
results from having part of the requested file already available in the
local cache. The third factor is a \emph{global} gain that arises from
using the caches to create simultaneous coded-multicasting
opportunities. See Example~\ref{eg:illustration} in
Section~\ref{sec:main_proposed} for an illustration of the operational
meaning of these three factors.

\subsection{Comparison with Decentralized Uncoded Caching Scheme} 
\label{sec:performance_decentralized}

It is instructive to examine the performance of the proposed
decentralized coded caching scheme (Algorithm~\ref{alg:1}) for large and
small values of cache size $M$.  For simplicity, we focus here on the
most relevant case $N \geq K$, i.e., the number of files is at least as
large as the number of users, so that the rate $R_D(M)$ of
Algorithm~\ref{alg:1} is given by Remark~\ref{rem:thm}. 

As a baseline, we compare the result with the uncoded caching scheme,
introduced in Example~\ref{eg:conventional} in
Section~\ref{sec:problem}. This is the best previously known algorithm
with a decentralized placement phase. For $N\geq K$, the uncoded
scheme achieves the rate 
\begin{equation}
    \label{eq:conventional2}
    R_U(M) = K\cdot(1-M/N),
\end{equation}
which is linear with slope of $-K/N$ throughout the entire range of $M$.

\subsubsection{Small $M$}

For small cache size $M\in[0,N/K]$, the rate achieved by
Algorithm~\ref{alg:1} behaves approximately\footnote{More precisely,
$R_D(M) = K - \frac{K(K+1)}{2N}M+O(M^2)$, and by analyzing the constant
in the $O(M^2)$ expression it can be shown that this is a good
approximation in the regime $M\in[0,N/K]$.} as
\begin{equation}
    \label{eq:proposed_small}
    R_D(M) \approx K\cdot \Bigl( 1-\frac{KM}{2N} \Bigr).
\end{equation}
In this regime, $R_D(M)$ scales approximately linearly with the memory
size $M$ with slope $-K^2/(2N)$: increasing $M$ by one decreases the
rate by approximately $K^2/(2N)$. This is illustrated in
Fig.~\ref{fig:regimes}.

Comparing~\eqref{eq:conventional2}
and~\eqref{eq:proposed_small}, we have the following
observations: 
\begin{itemize}
    \item \emph{Order-$K$ Improvement in Slope:} 
        The slope of $R_D(M)$ around $M=0$ is approximately $K/2$
        times steeper than the slope of $R_U(M)$.  Thus, the reduction
        in rate as a function of $M$ is approximately $K/2$ times faster
        for Algorithm~\ref{alg:1} than for the uncoded scheme.  In
        other words, for small $M$ the scheme proposed here makes
        approximately $K/2$ times better use of the cache resources: an
        improvement on the order of the number of users in the system.
        This behavior is clearly visible in Fig.~\ref{fig:intro} in
        Section~\ref{sec:intro}.   
    \item \emph{Virtual Shared Cache:} Consider a hypothetical setting
        in which the $K$ cache memories are collocated and shared among
        all $K$ users. In this hypothetical system, arising from
        allowing complete cooperation among the $K$ users, it is easy to
        see that the optimal rate is $K\cdot(1-KM/N)$.  Comparing this
        to~\eqref{eq:proposed_small}, we see that, up to a factor $2$,
        the proposed decentralized coded caching scheme achieves the
        same order behavior.  Therefore, this scheme is essentially able
        to create a single virtually shared cache, even though the
        caches are isolated without any cooperation between them.
\end{itemize}

\subsubsection{Large $M$}             

\begin{figure}[htbp]
    \centering
    \hspace{-0.8cm}\includegraphics{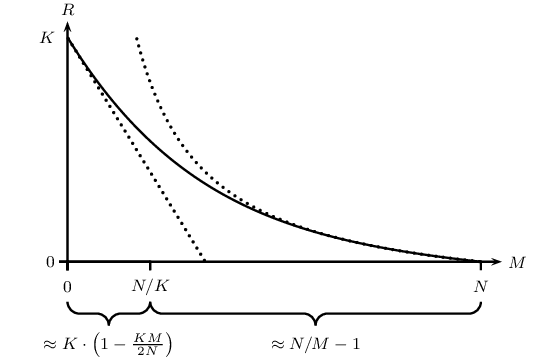}
    \caption{Memory-rate tradeoff $R_D(M)$ achieved by
    Algorithm~\ref{alg:1} for $N=100$ files and $K=5$ users (see
    Theorem~\ref{thm:achievable1}). The function $R_D(M)$ behaves
    approximately linearly for $M\in[0,N/K]$ and behaves approximately
    as $N/M-1$ for $M\in[N/K,N]$ (both
    approximations are indicated by dotted curves).} 
    \label{fig:regimes}
\end{figure}

On the other hand, for $M\in[N/K,N]$ we can approximate\footnote{Since
$(1-M/N)^K \leq (1-1/K)^K \leq 1/e$, we have $R_D(M) = \Theta(N/M-1)$ in
the regime $M\in[N/K,N]$, and the pre-constant in the order notation
converges to $1$ as $M\to N$.}
\begin{equation}
    \label{eq:proposed_large}
    R_D(M) \approx N/M-1.
\end{equation}
In this regime, the rate achieved by Algorithm~\ref{alg:1} scales
approximately inversely with the memory size: doubling $M$ approximately
halves the rate.  This is again illustrated in Fig.~\ref{fig:regimes}.

Comparing~\eqref{eq:conventional2} and~\eqref{eq:proposed_large}, we
have the following observation: 
\begin{itemize}
    \item \emph{Order-$K$ Improvement in Rate:} In this regime, the rate
        of the proposed decentralized coded scheme can be up to a factor
        $K$ smaller than the uncoded scheme: again an improvement on the
        order of the number of users in the system. This behavior is
        again clearly visible in Fig.~\ref{fig:intro} in
        Section~\ref{sec:intro}.
\end{itemize}

\subsection{Comparison with Centralized Coded Caching Scheme} 
\label{sec:performance_centralized}

We have compared the performance of the proposed decentralized coded
caching scheme to the uncoded caching scheme, which is the best
previously known \emph{decentralized} algorithm for this setting.  We
now compare the decentralized coded caching scheme to the rate
achievable by \emph{centralized} caching schemes. We start with an
information-theoretic lower bound on the rate of \emph{any} centralized
caching scheme. We then consider the rate of the best known centralized
caching scheme recently introduced in~\cite{maddah-ali12a}.

\begin{theorem}
    \label{thm:comparison_converse}
    Let $R_D(M)$ be the rate of the decentralized coded caching scheme
    given in Algorithm~\ref{alg:1}, and let $R^\star(M)$ be the rate of
    the optimal centralized caching scheme. For any number of files $N$
    and number of users $K$ and  for any $M\in[0,N]$, we have
    \begin{equation*}
        \frac{R_D(M)}{R^\star(M)} \leq 12.
    \end{equation*}
\end{theorem}

The proof of Theorem~\ref{thm:comparison_converse}, presented in
Appendix~\ref{sec:proofs_comparison_converse}, uses an
information-theoretic argument to lower bound the rate of the optimal
scheme $R^\star(M)$. As a result, Theorem~\ref{thm:comparison_converse}
implies that \emph{no} scheme (centralized, decentralized, with linear
caching, nonlinear caching, \ldots) regardless of is computational
complexity can improve by more than a constant factor upon the efficient
decentralized caching scheme given by Algorithm~\ref{alg:1} presented in
this paper. 

\begin{remark}[\emph{Optimality of Uncoded Prefetching and Linearly-Coded Delivery}]
    Theorem~\ref{thm:comparison_converse} implies that uncoded caching
    in the placement phase combined with greedy linear coding in the
    delivery phase is sufficient to achieve a rate within a constant
    factor of the optimum. 
\end{remark}

We now compare the rate $R_D(M)$ of decentralized coded caching to the
rate $R_C(M)$ of the best known centralized coded caching scheme.
In~\cite[Theorem~2]{maddah-ali12a}, $R_C(M)$ is given by 
\begin{equation*}
    R_C(M) \defeq K\cdot (1-M/N) \cdot \min\biggl\{ 
        \frac{1}{1+KM/N}, \, \frac{N}{K} 
    \biggr\}
\end{equation*}
for $M\in\tfrac{N}{K}\{0,1,\dots,K\}$, and the lower convex envelope of
these points for all remaining values of $M\in[0,N]$.
Fig.~\ref{fig:intro} in Section~\ref{sec:intro} compares the performance
$R_C(M)$ of this centralized coded caching scheme to the performance
$R_D(M)$ of the decentralized coded caching scheme proposed here. As can
be seen from the figure, the centralized and decentralized caching
algorithms are very close in performance. Thus, there is only a small
price to be paid for decentralization. Indeed, we have the following
corollary to Theorem~\ref{thm:comparison_converse}.

\begin{corollary}
    \label{thm:comparison_centralized}
    Let $R_D(M)$ be the rate of the decentralized coded caching scheme
    given in Algorithm~\ref{alg:1}, and let $R_C(M)$ be the rate of the
    centralized coded caching scheme from~\cite{maddah-ali12a}. For any
    number of files $N$ and number of users $K$ and  for any
    $M\in[0,N]$, we have
    \begin{equation*}
        \frac{R_D(M)}{R_C(M)} \leq 12.
    \end{equation*}
\end{corollary}

Corollary~\ref{thm:comparison_centralized} shows that the rate achieved
by the decentralized coded caching scheme given by Algorithm~\ref{alg:1}
is at most a factor $12$ worse than the one of the best known
centralized algorithm from \cite{maddah-ali12a}. This bound can be
tightened numerically to
\begin{equation*}
    \frac{R_D(M)}{R_C(M)} \leq 1.6
\end{equation*}
for all values of $K$, $N$, and $M$. Hence, the rate of the
decentralized caching scheme proposed here is indeed quite close to the
rate of the best known centralized caching scheme.

It is instructive to understand why the decentralized scheme performs
close to the centralized one. In the centralized scheme, content is
placed in the placement phase such that in the delivery phase every
message is useful for exactly $1+KM/N$ users. In the decentralized
scheme, we cannot control the placement phase as accurately. However,
perhaps surprisingly, the number of messages that are useful for about
$1+KM/N$ users is nevertheless the dominant term in the overall rate of
the decentralized scheme.
    
\begin{figure}[htbp]
    \centering
    \hspace{1.5cm}\includegraphics{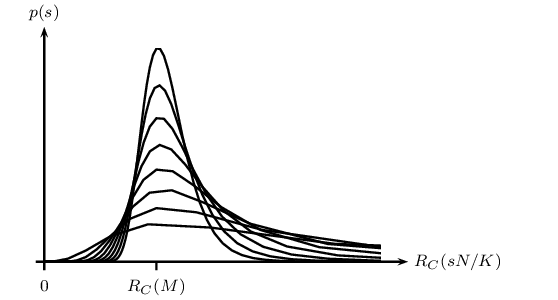}
    \caption{Concentration of the rate terms in the convex
        combination~\eqref{eq:comb} expressing the rate of the decentralized
        coded caching scheme $R_D(M)$ around the rate $R_C(M)$ of the
        centralized coded caching scheme. The curves are for different
        values of $N \in\{2^3,2^4,\dots,2^{10}\}$ with $K=N$ and
        $M=\sqrt{N}$. Each curve depicts $p(s)$ versus
        $R_C(sN/K)$ parametrized by $s\in\{0,1,\dots, K\}$.} 
    \label{fig:concentration}
\end{figure}

More precisely, from the proof of Theorem~\ref{thm:achievable1} in
Appendix~\ref{sec:proofs_achievable1}, we can write the rate
$R_D(M)$ of the decentralized coded scheme as a convex combination
of the rate $R_C(M)$ of the centralized coded scheme:
\begin{equation}
    \label{eq:comb}
    R_D(M) = \sum_{s=0}^{K} R_C(sN/K) p(s)
\end{equation}
with $p(s) \geq 0$ and $\sum_{s=0}^{K}p(s) = 1$. The dominant term
in this sum is $R_C(M)$ occurring at $s=KM/N$, as is illustrated in 
Fig.~\ref{fig:concentration}. This observation explains why the
centralized and the decentralized schemes have approximately the
same rate.

\section{Extensions}
\label{sec:extensions}

In this section, we extend the results presented so far to some
important cases arising in practical systems. In particular, we show how
to handle networks with tree topologies in
Section~\ref{sec:extensions_tree}, caches shared by several users in
Section~\ref{sec:extensions_shared}, and users with asynchronous
requests in Section~\ref{sec:extensions_synchronization}.

\subsection{Tree Networks}
\label{sec:extensions_tree}

\begin{figure}[htbp]
    \centering
    \includegraphics{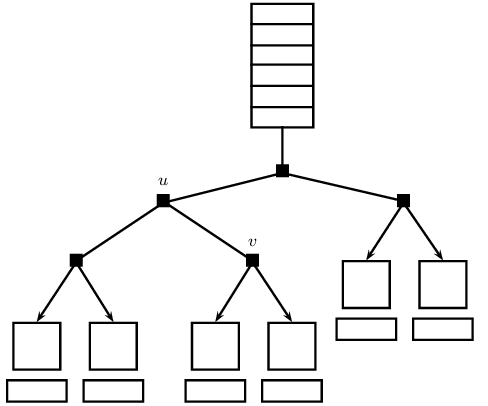}
    \caption{Network with tree structure. A server containing $N$ files
    of size $F$ bits each is connected through a tree-structured
    network to $K$ users each with a cache of size $MF$ bits.
    Internal nodes of the tree represent routers. In this figure,
    $N=K=6$, and $M=1$. The proposed placement and delivery
    procedures together with a routing algorithm achieves the
    order-optimal rate over every link $(u,v)$ of the network.} 
    \label{fig:tree}
\end{figure}

The basic problem setting considered so far considers users connected to
the server through a single shared bottleneck link. We showed that the
rate of our proposed algorithm over the shared link is within a constant
factor of the optimum. Here we extend this result to more general
networks with tree structure (see Fig.~\ref{fig:tree}). 

Consider a directed tree network, oriented from the root to the leaves.
The server is located at the root of the tree, and users with their
caches are located at the leaves.  Each internal node of the network
represents a router. The router decides what to transmit over each of
its outgoing links as a function of what it received over its single
incoming link from its parent.

We again assume that the system operates in  two phases. In the
placement phase, the caches are populated without knowledge of users'
future demands. In the delivery phase, the users reveal their requests,
and the server has to satisfy these demands exploiting the cached
content. 

For this network, we propose the following caching and routing
procedures. For the placement phase, we use the same placement procedure as
in Algorithm~\ref{alg:1}. For the delivery phase, we use the two
delivery procedures detailed in Algorithm~\ref{alg:1}, but with the
simplified decision rule explained in Remark~\ref{Remark:M}. In other
words, if $M>1$, the server creates coded packets according to the first
delivery  procedure. If $M \leq 1$, it uses the second delivery
procedure and the server creates linear combinations of the bits of each
file without coding across different files. 

It remains to describe the operations of the routers at the internal
nodes of the tree. Each router operates according to the following
simple rule. The router at node $u$ forwards a coded message over link
$(u,v)$ if and only if that coded message is directly useful to at least
one of the descendant leaves of node $v$. To be precise, let us assume
that $M>1$ so that the server uses the first delivery procedure. Thus
for each subset $\mc{S} \subset [K]$ of users, the server creates the
coded message $\oplus_{k\in\mc{S}} V_{k,\mc{S}\setminus\{k\}}$. This
coded message is useful for all users in $\mc{S}$ and is completely
useless for the remaining users. The router located at node $u$ forwards
this coded message $\oplus_{k\in\mc{S}} V_{k,\mc{S}\setminus\{k\}}$ over
the link $(u,v)$, if at least one of the descendants of $v$ (including
$v$ itself if it is a leaf) is in the set $\mc{S}$. A similar routing
procedure is used for $M \leq 1$. 

The performance of this scheme is analyzed in
Appendix~\ref{sec:proofs_tree}. We show there that, for $M > 1$, the
rate of this scheme over the link $(u,v)$ is equal to 
\begin{align}
    \label{eq:rate-tree}
    K_v \cdot (1-M/N) \cdot \frac{N}{K_v M}\bigl(1-(1-M/N)^{K_v}\bigr),
\end{align}
where $K_{v}$ is the number of descendant leaves of node $v$. For $M \leq
1$, it is easy to see that the rate over the link $(u,v)$ is equal to
\begin{align}
    K_v \cdot (1-M/N) \cdot \min\{ 1, N/K_{v}\} .
\end{align}

The rate over every link in the tree network can be shown to be within a
constant factor of optimal. To prove approximate optimality for edge
$(u,v)$, we consider the subtree rooted at $v$ together with $u$ and the
edge $(u,v)$. We can then use the same bound used in
Theorem~\ref{thm:comparison_converse} over this subtree, treating
$(u,v)$ as the shared bottleneck link. 

\begin{remark}[\emph{Universality and Separation of Caching and Routing}]
    This result shows that, for tree-structured networks, caching and
    routing can be performed separately with at most a constant factor
    loss in performance compared to the optimal joint scheme. This means
    that the proposed placement and delivery procedures are universal in
    the sense that they do not depend on the topology of the tree network
    connecting the server to the caches at the leaves. 
\end{remark}

\begin{example}[\emph{Universality}]
    \label{eg:universal}
    Consider $K$ users connected to a server through $K$ orthogonal
    links (i.e., no shared links). For this topology the optimal rate
    over each link can be achieved without coding.  However, it is easy
    to see that the proposed universal scheme achieves the same optimal
    rate. Thus, depending on the network topology, we may be able to
    develop simpler schemes, but the performance will be the same up to
    a constant factor as the proposed universal scheme. 
\end{example}

\begin{example}[\emph{Rate over Private Links}]
    \label{sec:private}
    Consider the original scenario of users sharing a single bottleneck
    link. As an example, assume we have $N=2$ files and $K=2$ users as
    described in Example~\ref{eg:illustration} in
    Section~\ref{sec:main_proposed}. Observe that a user does not need
    all messages sent by the server over the shared link in order to
    recover its requested file. For example, in order to recover file
    $A$, user one only needs $A_{2} \oplus B_{1}$ and $A_\emptyset$.
    Thus, if a router is located right where the shared link splits into
    the two private links, it can forward only these two messages over
    the private link to user one. By the analysis in this section, the
    resulting normalized rate over the private link to user one is then 
    \begin{equation*}
        (M/2)(1-M/2) + (1-M/2)^{2} = 1-M/N.
    \end{equation*}

    We note that in the uncoded scheme the rate over each private link
    is also $1-M/N$. Hence, we see that by proper routing the rate over
    the private links for both the coded as well as the uncoded schemes
    are the same. The reduction of rate over the shared link achieved by
    coding does therefore not result in an increase of rate over the
    private links. This conclusion holds also for general values of $N$,
    $K$, and $M$.
\end{example}

In this section, we have only considered tree networks with caches at
the leaves. The general scenario, in which caches are also present at
internal nodes of the tree, is more challenging and is analyzed in
follow-up work~\cite{karamchandani14}.

\subsection{Shared Caches}
\label{sec:extensions_shared}

The problem setting considered throughout this paper assumes that
each user has access to a private cache. In this example, we
evaluate the gain of shared caches. This situation arises when the
cache memory is located close to but not directly at the users.

\begin{figure}[htbp]
    \centering
    \includegraphics{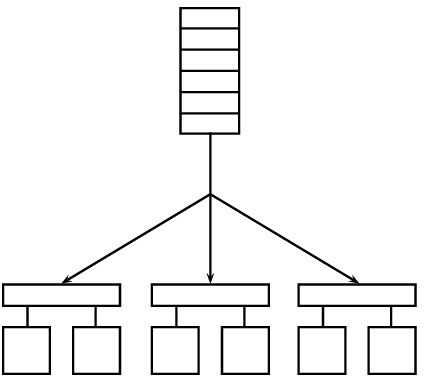}
    \caption{Users with shared caches. A server containing $N$ files of
    size $F$ bits each is connected through a shared link to $K/L$
    caches each of size $LMF$ bits. Each cache is shared among $L$ users. 
    In the figure, $K=N=6$ and $L=2$.}
    \label{fig:shared}
\end{figure}

We consider a system with  $K$ users partitioned into subsets, where
users within the same subset share a common cache (see
Fig.~\ref{fig:shared}). For simplicity, we assume that these subsets
have equal size of $L$ users, where $L$ is a positive integer dividing
$K$. We also assume that the number of files $N$ is greater than the
number of users $K$. To keep the total amount of cache memory in the
system constant, we assume that each of the shared caches has size $LMF$
bits. 

We can operate this system as follows. Define $K/L$ super users, one for
each subset of $L$ users. Run the placement procedure of
Algorithm~\ref{alg:1} for these $K/L$ users with cache size $LMF$.  In
the delivery phase, treat the (up to) $L$ files requested by the users
in the same subset as a single super file of size $LF$.  Applying
Theorem~\ref{thm:achievable1} to this setting yields an achievable rate
of\footnote{This can be derived from $R_D(M)$ in
Theorem~\ref{thm:achievable1} by replacing $M$ by $LM$ (since each
cache has now size $LMF$ instead of $MF$), replacing $K$ by $K/L$
(since there are $K/L$ super users), and multiplying the result by
an extra factor of $L$ (since each super file is $L$ times the size of a
normal file).}
\begin{equation*}
    R_D^L(M)=  K \cdot(1-LM/N)\cdot\frac{N}{KM}\bigl(1-(1-LM/N)^{K/L}\bigr).
\end{equation*}

Let us again consider the regimes of small and large values of $M$
of $R_D^L(M)$. For $M\in[0,N/K]$, we have
\begin{equation*}
    R_D^L(M) \approx K\cdot\Bigl(1-\frac{K+L}{2N}M\Bigr).
\end{equation*}
Comparing this to the small-$M$ approximation~\eqref{eq:proposed_small}
of $R_D(M)$ (for a system with private caches), we see that
\begin{equation*}
    R_D^L(M) \approx R_D(M),
\end{equation*}
i.e., there is only a small effect on the achievable rate from sharing a
cache. This should not come as a surprise, since we have already seen in
Section~\ref{sec:performance_decentralized} that, for small $M$, $R_D(M)$
behaves almost like a system in which all $K$ caches are combined.
Hence, there is no sizable gain to be achieved by having collaboration
among caches in this regime.

Consider then the regime $M> N/K$. Here, we have
\begin{equation*}
    R_D^L(M) \approx N/M-L = K\cdot(1-LM/N)\cdot\frac{N}{KM} 
\end{equation*}
and from~\eqref{eq:proposed_large}
\begin{equation*}
    R_D(M) \approx N/M-1 = K\cdot(1-M/N)\cdot\frac{N}{KM}.
\end{equation*}
The difference between the two approximations is only in the second
factor. We recall that this second factor represents the caching
gain due to making part of the files available locally. Quite
naturally, this part of the caching gain improves through cache
sharing, as a larger fraction of each file can be stored locally.

\subsection{Asynchronous User Requests}
\label{sec:extensions_synchronization}

Up to this point, we have assumed that in the delivery phase all users
reveal their requests simultaneously, i.e., that the users are perfectly
synchronized. In practice, however, users reveal their requests at
different times. In this example, we show that the proposed algorithm
can be modified to handle such asynchronous user requests. 

We explain the main idea with an example. Consider a system with
$N=3$ files $A, B, C$, and $K=3$ users.  We split each file into
$J$ consecutive segments, e.g., $A=(A^{(1)}, \dots, A^{(J)})$
and similarly for $B$ and $C$.  Here $J$ is a positive integer
selected depending on the maximum tolerable delay, as will be
explained later. To be specific, we choose $J=4$ in this
example. 

\begin{figure}[htbp]
    \centering
    \hspace{-0.2cm}\includegraphics{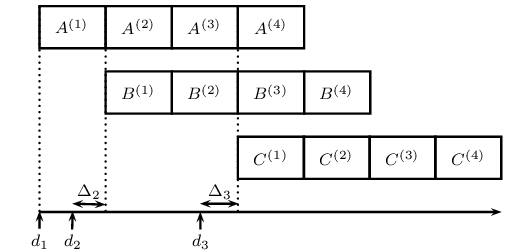}
    \caption{The proposed scheme over segmented files can be used
        to handle asynchronous user requests. In this example, each file is
        split into four segments. Users two and three are served with a
    delay of $\Delta_2$ and $\Delta_3$, respectively.} 
    \label{fig:segments}
\end{figure}

In the placement phase, we simply treat each segment as a file.  We
apply the placement procedure of Algorithm~\ref{alg:1}. For the delivery
phase, consider an initial request $d_1$ from user one, say for file
$A$. The server responds by starting delivery of the first segment
$A^{(1)}$ of file $A$. Meanwhile, assume that user two requests file
$d_2$, say $B$, as shown in Fig.~\ref{fig:segments}.  The server puts
the request of user two on hold, and completes the delivery of $A^{(1)}$
for user one. It then starts to deliver the second segment $A^{(2)}$ of
$A$ and the first segment $B^{(1)}$ of $B$ using the delivery procedure
of Algorithm~\ref{alg:1} for two users. Delivery of the next segments
$A^{(3)}$ and $B^{(2)}$ is handled similarly. Assume that at this point
user three requests file $d_3$, say $C$, as shown in
Fig.~\ref{fig:segments}.  After completing the current delivery phase,
the server reacts to this request by delivering $A^{(4)}$, $B^{(3)}$,
and $C^{(1)}$ to users one, two, and three, respectively, using the
delivery procedure of Algorithm~\ref{alg:1} for three users. The process
continues in the same manner as depicted in Fig.~\ref{fig:segments}. 

We note that users two and three experience delays of $\Delta_2$ and
$\Delta_3$ as shown in the figure. The maximum delay depends on the size
of the segments. Therefore, segment size, or equivalently the value of
$J$, can be adjusted to ensure that this delay is tolerable (while
keeping the segment size large enough to ensure that the law of large
number applies as discussed in Example~\ref{eg:illustration} in
Section~\ref{sec:main_proposed}).

We point out that the number of effective users in the system varies
throughout the delivery phase. Due to its decentralized nature, the
proposed caching algorithm is close to optimal for any value of
users as discussed in Remark~\ref{rem:knownK}. This is instrumental
for the segmentation approach just discussed to be efficient.

\section{Discussion}
\label{sec:discussions}

\subsection{Connection to Index and Network Coding}

The caching problem considered in this paper is connected to the index
coding problem~\cite{birk06, bar-yossef11} (or,
equivalently~\cite{effros12}, to the network coding
problem~\cite{ahlswede00}). In the index coding problem, we are given a
set of $K$ users and a set of $N$ files. Each of the users has access to
a fixed subset of those files and requests another fixed subset of the
files. The goal is to broadcast the minimum number of bits to the
$K$ users in order to satisfy all user requests.

From the above description, we see that for \emph{fixed} content
placement and for \emph{fixed} user demands, the caching problem
considered in this paper induces an index coding problem in the delivery
phase. Since there are $N^K$ possible user demands, the delivery phase
of the caching problem can thus be interpreted as an exponential number
of parallel index coding subproblems. To complicate matters, the index
coding problem itself does not admit a closed-form solution and is known
to be computationally intractable in general~\cite{langberg11}.

One contribution of this paper is thus the design of the placement phase
such that all these exponentially many index coding problems
simultaneously have an efficient and closed-form solution.

\subsection{Caching Random Linear Combinations is Inefficient}

Caching random linear combinations of file segments is a popular
prefetching scheme. In this example, we argue that in some scenarios
this form of caching can be quite inefficient.

To be precise, let us focus on a specific scenario with $K$ users and
$N=K$ files, where each user has sufficient cache memory to store half
of the files, i.e. $M=N/2=K/2$. According to
Theorem~\ref{thm:achievable1}, Algorithm~\ref{alg:1} achieves a rate of
less than one, i.e., $R_D(M)\leq 1$.

On the other hand, the rate achieved by caching of random linear
combinations can be shown to be at least $K/4$, which is significantly
larger than $R_D(M)$ for large number of users $K$. Indeed, assume that
user one requests file $A$. Recall that each user has cached $F/2$
random linear combinations of the bits of file $A$. With high probability, these
random linear combinations span a $F/2$-dimensional space at each user
and the subspaces of different users do not overlap. For example,
consider users two and three. As a consequence of this lack of overlap,
these two users do not have access to a shared part of the file $A$.
This implies that, in the delivery phase, the server cannot form a
linear combination that is simultaneously useful for three users. In
other words, the server can form messages that are at most useful
simultaneously for up to two users. A short calculation reveals that
then the server has to send at least $FK/4$ bits over the shared link.

This inefficiency of caching random linear combinations can be
interpreted as follows. The placement phase follows  two competing
objectives: The first objective is to spread the available content as
much as possible over the different caches. The second objective, is to
ensure maximum overlap among different caches. The system performance is
optimized if the right balance between these two objectives is struck.
Caching random linear combinations maximizes the spreading of content
over the available caches, but provides minimal overlap among them. At
the other extreme, the uncoded caching scheme maximizes the overlap, but
provides only minimal spreading of the content. As a consequence, both
of these schemes can be highly suboptimal.

\subsection{Worst-Case Demands}

Our problem formulation focuses on worst-case requests. In some
situation, this is the correct figure of merit. For example, in a
wireless scenario, whenever the delivery rate required for a request
exceeds the available link bandwidth, the system will be in outage,
degrading user experience. In other situations, for example a wireline
scenario, excess rates might only incur a small additional cost and
hence might be acceptable. In such cases, the average rate is the right
metric, especially when files have different popularities.  This is
discussed further in~\cite{niesen13}.

\subsection{Online Coded Caching}

In practical scenarios, the set of popular files is time varying.  To
keep the caching algorithm efficient, the cache contents have to be
dynamically updated to track this variation. A popular scheme to update
the caches is to evict the least-recently used (LRU) file from the caches
and replace it with a newly requested one. This LRU eviction scheme is
known to be approximately optimal for systems with a single
cache~\cite{sleator85}. However, it is not efficient for networks with
multiple caches as considered here. This problem of online caching with
several caches is investigated in~\cite{pedarsani13}. The decentralized
Algorithm~\ref{alg:1} presented in this paper turns out to be a crucial
ingredient of the suggested online coded caching algorithm
in~\cite{pedarsani13}.

\appendices

\section{Proof of Theorem~\ref{thm:achievable1}}
\label{sec:proofs_achievable1}

We first prove correctness. Note that, since there are a total of $N$
files, the operations in Line~\ref{alg:1_cache} of Algorithm~\ref{alg:1}
satisfies the memory constraint of $MF$ bits at each user. Hence the
placement phase of Algorithm~\ref{alg:1} is correct.

For the delivery phase, assume the server uses the first delivery
procedure, and consider a bit in the file requested by user $k$. If
this bit is already cached at user $k$, it does not need to be sent
by the server. Assume then that it is cached at some (possibly
empty) set $\mc{T}$ of users with $k\notin\mc{T}$. Consider the set
$\mc{S}=\mc{T}\cup\{k\}$ in Line~\ref{alg:1_Sloop}. By definition,
the bit under consideration is contained in
\(V_{k,\mc{S}\setminus\{k\}}\), and as a consequence, it is included
in the sum sent by the server in Line~\ref{alg:1_send}. Since \(k\in
\mc{S}\setminus\{\tilde{k}\}\) for every other $\tilde{k}\in\mc{S}$,
user $k$ has access to all bits in
\(V_{\tilde{k},\mc{S}\setminus\{\tilde{k}\}}\) from its own cache.
Hence, user $k$ is able to recover the requested bit from this sum.
This shows that the first delivery procedure is correct.

The second delivery procedure is correct as well since the server
sends in Line~\ref{alg:1_send2} enough linear combinations of every
file for all users to successfully decode.  This shows that the
delivery phase of Algorithm~\ref{alg:1} is correct.

It remains to compute the rate. We start with the analysis of the second
delivery procedure. If $N \leq K$, then in the worst case there is at
least one user requesting every file. Consider then all users requesting
file $n$. Recall that each user requesting this file already has $FM/N$
of its bits cached locally by the operation of the placement phase. An
elementary analysis reveals that with high probability for $F$ large
enough at most
\begin{equation*}
    F(1-M/N)+o(F)
\end{equation*}
random linear combinations need to be sent in Line~\ref{alg:1_send} for
all those users to be able to decode. We will assume that the file size
$F$ is large and ignore the $o(F)$ term in the following. Since this
needs to be done for all $N$ files, the normalized rate in the delivery
phase is
\begin{equation*}
    (1-M/N)N.
\end{equation*}

If $N > K$, then there are at most $K$ different files that are
requested. The same analysis yields a normalized  rate of
\begin{equation*}
    (1-M/N)K.
\end{equation*}
Thus, the second procedure has a normalized rate of 
\begin{align}
    \label{eq:second}
    R(M) & = (1-M/N)\min\{K,N\} \notag\\
    & = K\cdot(1-M/N)\cdot\min\{1,N/K\}
\end{align}
for $M\in(0,N]$. 

We continue with the analysis of the first delivery procedure.
Consider a particular bit in one of the files, say file $n$. Since
the choice of subsets is uniform, by symmetry this bit has
probability
\begin{equation*}
q \defeq M/N \in (0,1]
\end{equation*}
of being in the cache of any fixed user. Consider now a fixed subset
of $t$ out of the $K$ users. The probability that this bit is cached
at exactly those $t$ users is
\begin{equation*}
    q^t(1-q)^{K-t}.
\end{equation*}
Hence the expected number of bits of file $n$ that are cached at
exactly those $t$ users is
\begin{equation*}
    Fq^t(1-q)^{K-t}.
\end{equation*}
In particular, the expected size of \(V_{k,\mc{S}\setminus\{k\}}\)
with $\card{\mc{S}}=s$ is
\begin{equation*}
    Fq^{s-1}(1-q)^{K-s+1}.
\end{equation*}
Moreover, for $F$ large enough the actual
realization of the random number of bits in \(V_{k,\mc{S}\setminus\{k\}}\)
is in the interval 
\begin{equation*}
    Fq^{s-1}(1-q)^{K-s+1}\pm o(F)
\end{equation*}
with high probability. For ease of exposition, we will again ignore the
$o(F)$ term in the following.

Consider a fixed value of $s$ in Line~\ref{alg:1_sloop} and a fixed
subset $\mc{S}$ of cardinality $s$ in Line~\ref{alg:1_Sloop}. In
Line~\ref{alg:1_send}, the server sends
\begin{equation*}
    \max_{k\in\mc{S}}\card{V_{k,\mc{S}\setminus\{k\}}}
    = Fq^{s-1}(1-q)^{K-s+1}
\end{equation*}
bits. Since there are $\binom{K}{s}$ subsets $\mc{S}$ of cardinality
$s$, the loop starting in Line~\ref{alg:1_Sloop} generates 
\begin{equation*}
    \binom{K}{s}Fq^{s-1}(1-q)^{K-s+1}
\end{equation*}
bits. Summing over all values of $s$ yields a total of
\begin{align*}
    R_D(M)F
    & = F\sum_{s=1}^K \binom{K}{s}q^{s-1}(1-q)^{K-s+1} \\
    & = F\frac{1-q}{q}\Bigg(\sum_{s=0}^K \binom{K}{s}q^{s}(1-q)^{K-s}-(1-q)^K\Bigg) \\
    & = F\frac{1-q}{q}\big(1-(1-q)^K\big)
\end{align*}
bits being sent over the shared link. Substituting the definition of
$q=M/N$ yields a rate of the first delivery procedure of
\begin{align} 
    \label{eq:first}
    R_D(M) 
    & = (N/M-1)\bigl(1-(1-M/N)^K\bigr) \nonumber\\
    & = K\cdot(1-M/N)\cdot\frac{N}{KM}\bigl(1-(1-M/N)^K\bigr)
\end{align}
for $M\in(0,N]$.

Since the server uses the better of the two delivery
procedures, \eqref{eq:second} and~\eqref{eq:first} show that
Algorithm~\ref{alg:1} achieves a rate of
\begin{align*}
    R_D(M)
    = K\cdot(1-M/N)\cdot\min\biggl\{
        \frac{N}{KM}\bigl(1-(1-M/N)^K\bigr), 1, \frac{N}{K} 
    \biggr\}.
\end{align*}
Using that 
\begin{equation*}
    (1-M/N)^K \geq 1-KM/N,
\end{equation*}
this can be simplified to 
\begin{equation*}
    K\cdot(1-M/N)\cdot\min\biggl\{
        \frac{N}{KM}\bigl(1-(1-M/N)^K\bigr), \frac{N}{K} 
    \biggr\},
\end{equation*}
concluding the proof. \hfill\qedhere

\section{Proof of Theorem~\ref{thm:comparison_converse}}
\label{sec:proofs_comparison_converse}

Recall from Theorem~\ref{thm:achievable1} that
\begin{align*}
    R_D(M) 
    = K\cdot(1-M/N)\cdot\min\biggl\{ 
        \frac{N}{KM}\bigl(1-(1-M/N)^K\bigr), \frac{N}{K}
    \biggr\}.
\end{align*}
Using that
\begin{align*}
    (1-M/N)^K & \geq 0 \\
    \shortintertext{and} \\
    (1-M/N)^K & \geq 1-KM/N,
\end{align*}
this can be upper bounded as
\begin{align}
    \label{eq:converse1}
    R_D(M) \leq \min\bigl\{N/M-1,K(1-M/N), N(1-M/N)\}
\end{align}
for all $M\in[0,N]$. Moreover, we have from
\cite[Theorem~2]{maddah-ali12a}
\begin{equation}
    \label{eq:converse2}
    R^\star(M)
    \geq \max_{s\in \{1, \dots, \min\{N, K\} \}}
    \Bigl(s-\frac{s}{\floor{N/s}}M\Bigr).
\end{equation}

We will treat the cases $0 \leq \min\{N,K\} \leq 12$ and $ \min\{N,K\}  \geq 13$
separately. Assume first that $0 \leq  \min\{N,K\}  \leq 12$.
By~\eqref{eq:converse1}, 
\begin{equation*}
    R_D(M) \leq  \min\{N,K\} (1-M/N) \leq 12(1-M/N),
\end{equation*}
and by~\eqref{eq:converse2} with $s=1$,
\begin{equation*}
    R^\star(M) \geq 1-M/N.
\end{equation*}
Hence
\begin{equation}
    \label{eq:converse3}
    \frac{R_D(M)}{R^\star(M)} \leq 12
\end{equation}
for $0 \leq  \min\{N,K\}  \leq 12$.

Assume in the following that $ \min\{N,K\}  \geq 13$. We consider the
cases 
\begin{equation*}
    M \in
    \begin{cases}
        \bigl[0,  \max \{1, N/K \} \bigr], \\
    \bigl(\max \{1, N/K \}, N/12\bigr], \\
\bigl(N/12, N\bigr],
    \end{cases}
\end{equation*}
separately. Assume first that $0\leq M \leq
\max \{1, N/K \}$.  By~\eqref{eq:converse1}, 
\begin{equation*}
    R_D(M) \leq \min\{N,K\}(1-M/N) \leq \min\{N,K\}.
\end{equation*}
On the other hand, by~\eqref{eq:converse2} with $s=\floor{\min\{N,K\}/4}$,
\begin{align*}
    \R^\star(M)
    & \geq s-\frac{s^2}{1-s/N}\frac{M}{N} \\
    & \geq \min\{N,K\}/4-1-\frac{(\min\{N,K\})^2/16}{1-\min\{N,K\}/(4N)}\frac{M}{N} \\
    & \stackrel{(a)}{\geq} \min\{N,K\}\Bigl(1/4-1/13-\frac{1/16}{1-1/4}\Bigr)\\
    & \geq  \min\{N,K\}/12,
\end{align*}
where in $(a)$ we have used that $ \min\{N,K\} \geq 13$ and $M \leq \max \{1, N/K \}$. Hence
\begin{equation}
    \label{eq:converse4}
    \frac{R_D(M)}{R^\star(M)} \leq 12
\end{equation}
for $\min{N,K} \geq 13$ and $0\leq M \leq \max \{1, N/K \}$.

Assume then that $\max \{1, N/K \}< M \leq N/12$. By~\eqref{eq:converse1},
\begin{equation*}
    R_D(M) \leq N/M-1 \leq N/M.
\end{equation*}
On the other hand, by~\eqref{eq:converse2} with $s=\floor{N/(4M)}$,
\begin{align*}
    R^\star(M)
    & \geq s-\frac{s^2}{1-s/N}\frac{M}{N} \\
    & \geq N/(4M)-1-\frac{N^2/(16M^2)}{1-1/(4M)}\frac{M}{N} \\
    & = \frac{N}{M}\Bigl(1/4-M/N-\frac{1/16}{1-1/(4M)}\Bigr) \\
    & \stackrel{(a)}{\geq} \frac{N}{M}\Bigl(1/4-1/12-\frac{1/16}{1-1/4}\Bigr) \\
    & = N/(12M),
\end{align*}
where in $(a)$ we have used that $M/N \leq 1/12$ and that $M >
\max \{1, N/K \} \geq 1$. Hence
\begin{equation}
    \label{eq:converse5}
    \frac{R_D(M)}{R^\star(M)} \leq 12
\end{equation}
for $\min\{N,K\} \geq 13$ and $\max \{1, N/K \} < M \leq N/12$.

Finally, assume that $N/12 < M \leq N$. By~\eqref{eq:converse1},
\begin{equation*}
    R_D(M) \leq N/M-1.
\end{equation*}
On the other hand, by~\eqref{eq:converse2} with $s=1$,
\begin{align*}
    R^\star(M) \geq 1-M/N.
\end{align*}
Hence,
\begin{align}
    \label{eq:converse6}
    \frac{R_D(M)}{R^\star(M)} 
    & \leq \frac{N/M-1}{1-M/N} \notag\\
    & = N/M \notag\\
    & \leq 12
\end{align}
for $\min\{N,K\} \geq 13$ and $N/12 < M \leq N$.

Combining~\eqref{eq:converse3}, \eqref{eq:converse4},
\eqref{eq:converse5}, and \eqref{eq:converse6} yields that
\begin{equation*}
    \frac{R_D(M)}{R^\star(M)} 
    \leq 12
\end{equation*}
for all $N$, $K$, and $0 \leq M\leq N$. \hfill\qedhere

\section{Proof of~\eqref{eq:rate-tree} in Section~\ref{sec:extensions_tree}}
\label{sec:proofs_tree}

As is shown in Appendix~\ref{sec:proofs_achievable1}, for $F$ large
enough the actual realization of the random number of bits in
\(V_{k,\mc{S}\setminus\{k\}}\) is in the interval 
\begin{equation*}
    Fq^{s-1}(1-q)^{K-s+1}\pm o(F)
\end{equation*}
with high probability, and where $q=M/N$. As before, we will again
ignore the $o(F)$ term in the following. 

Recall that only a subset of coded messages generated in
Line~\ref{alg:1_Sloop} in Algorithm~\ref{alg:1} pass through link
$(u,v)$, namely only those $\oplus_{k\in\mc{S}}
V_{k,\mc{S}\setminus\{k\}}$ for which the subset $\mc{S}$ has at least
one element among leave descendants of node $v$.  We split $\mc{S}$ into
$\mc{S}_1$ and $\mc{S}_2$ where $\mc{S}_1$ is the subset of descendant
leaves of node $v$, and $\mc{S}_2 \defeq \mc{S}\setminus\mc{S}_1$.
Denote the cardinalities  of $\mc{S}_1$ and $\mc{S}_2$ by $s_1$ and
$s_2$, respectively, so that $s = s_1+s_2$.  With this, only coded
messages with $s_1 \geq 1$ are forwarded over link $(u,v)$. The number
of bits sent over this link is then equal to
\begin{align*}
    F\sum_{s_1=1}^{K_v} \sum_{s_2=0}^{K-K_v} 
    & \binom{K_v}{s_1} \binom{K-K_v}{s_2}q^{s_1+s_2-1}(1-q)^{K-s_1-s_2+1} \\ 
    & = F \sum_{s_1=1}^{K_v}\binom{K_v}{s_1}  q^{s_1-1}(1-q)^{K_v-s_1+1}   
    \sum_{s_2=0}^{K-K_v}  \binom{K-K_v}{s_2}q^{s_2}(1-q)^{K-K_v-s_2} \\
    & = F \sum_{s_1=1}^{K_v}\binom{K_v}{s_1}  q^{s_1-1}(1-q)^{K_v-s_1+1} \\
    & = F\frac{1-q}{q}\big(1-(1-q)^{K_v}\big).
\end{align*}
Substituting $q=M/N$ yields the desired result.

\vfill

\end{document}